\let\orilabel\label
\documentclass[aps,pra,preprint,a4paper,showpacs,showkeys,superscriptaddress,nofootinbib]{revtex4-1}
\let\label\orilabel

\usepackage{latexsym}
\usepackage{amsmath,amssymb,mathrsfs}
\usepackage{mathtools}
\usepackage{graphicx}
\usepackage{subfigure}
\usepackage{xcolor}
\usepackage{physics}
\usepackage{cancel}
\usepackage[normalem]{ulem}
\usepackage{tikz}
\usepackage{multirow,tabularx}

\newcolumntype{C}[1]{>{\centering\arraybackslash}p{#1}}
\usepackage{hyperref}     
\hypersetup{colorlinks,%
  citecolor=blue,%
  linkcolor=cyan,%
}
\usepackage[titletoc]{appendix}
\usepackage{enumerate}
\graphicspath{{./Figures/}}
\usetikzlibrary{decorations.pathmorphing,decorations.pathreplacing,decorations.shapes}
\begin{document}


\title{The quasilocal energy and thermodynamic first law in accelerating AdS black holes}

\author{Wontae Kim}%
\email[]{wtkim@sogang.ac.kr}%
\affiliation{Department of Physics, Sogang University, Seoul, 04107,
	Republic of Korea}%

\author{Mungon Nam}%
\email[]{clrchr0909@sogang.ac.kr}%
\affiliation{Department of Physics, Sogang University, Seoul, 04107,
	Republic of Korea}%

\author{Sang-Heon Yi}%
\email[]{shyi@sogang.ac.kr}%
\affiliation{Department of Physics, Sogang University, Seoul, 04107,
	Republic of Korea}%
\date{\today}

\begin{abstract}
We scrutinize the conserved energy of an accelerating AdS black hole by employing the off-shell quasilocal formalism,
which amalgamates the  ADT formalism with the covariant phase space approach.
In the presence of conical singularities in the accelerating black hole, the energy expression is articulated through
the surface term derived from our formalism.
The essence of our analysis of the quasilocal energy
resides in the surface contributions  coming from the conical singularities as well as the conventional radial boundary.
Consequently, the resultant conserved quasilocal energy naturally conforms the thermodynamic first law for the black hole
without necessitating any augmentation of thermodynamic variables.
\end{abstract}
%



\maketitle


\raggedbottom

\section{Introduction}
\label{sec:introduction}
The thermodynamics of black holes provides deep insights into the quantum theory of gravity.
In the seminal works of Bekenstein~\cite{Bekenstein:1973ur,Bekenstein:1974ax} and Hawking~\cite{Hawking:1975vcx}, it was revealed that the area of a black hole and its surface gravity are correlated with the black hole's entropy and temperature.
Given that thermal properties of a physical system are closely intertwined with the statistical description of its microstates, black hole thermodynamics would elucidate the underlying microscopic degrees of freedom, potentially illuminating certain quantum aspects of gravity~\cite{Frolov:1993ym}.
Consequently, black hole thermodynamics has been investigated across numerous gravity models over the past decades.
In particular, recent studies have explored the thermodynamics of accelerating black holes~\cite{Dutta:2005iy,Bilal:2010ts,Appels:2016uha,Astorino:2016xiy,Astorino:2016ybm,Appels:2017xoe,Gregory:2017ogk,Anabalon:2018ydc,Anabalon:2018qfv,Golchin:2019hlg,Golchin:2020mkh,Ferrero:2020twa,Ball:2020vzo,Ball:2021xwt,Cassani:2021dwa,Arenas-Henriquez:2022www,Kim:2023ncn,Arenas-Henriquez:2023hur,Colombo:2024mts}, which are described by the C-metric as an exact solution to the Einstein field equations~\cite{Kinnersley:1970zw,Plebanski:1976gy,Dias:2002mi,Griffiths:2005qp,Griffiths:2009dfa}.
A notable feature of these black holes is the presence of at least one irremovable conical deficit angle along the azimuthal axis.
This conical singularity is responsible for the acceleration of the black hole, which may be understood by replacing it with an energy-momentum tensor corresponding to finite-width topological defects~\cite{Gregory:1995hd} or magnetic flux tubes~\cite{Dowker:1993bt}.

Exploring the thermodynamics of accelerating black holes in the presence of conical singularities remains challenging.
One main difficulty arises in defining the conserved energy of these black holes, which is complicated by their nontrivial asymptotic structure.
In particular, the conical singularities reach the conformal infinity of the black hole~\cite{Podolsky:2002nk,Griffiths:2006tk}, rendering the asymptotic structure topologically different from $\mathbb{R}\cross S^2$ due to the deficit angle.
Thus, this topological disparity disturbs the use of conformal regularization methods, developed initially by Ashtekar, Magnon, and Das (AMD)~\cite{Ashtekar:1984zz,Magnon:1985sc,Ashtekar:1999jx,Das:2000cu}, for determining the conserved energy of accelerating black holes.
Regarding the conserved energy of the black holes,
recent studies~\cite{Appels:2016uha,Appels:2017xoe,Gregory:2017ogk,Anabalon:2018ydc} also suggest
that string tensions associated with the conical singularities are required to be constant in order to satisfy the conventional form of the thermodynamic first law.

To effectively evade the complexities associated with the exotic asymptotic structure,
one can adopt a quasilocal formalism for the conserved energy.
The quasilocal approach, unlike global asymptotic methods, provides a robust framework for defining a conserved charge within a finite region of spacetime, thereby avoiding the difficulties associated with the asymptotic region.
For a comprehensive review of this approach, see Ref.~\cite{Szabados:2009eka} and references therein.
In particular, the quasilocal method developed in Ref.~\cite{Kim:2013zha}
derives the conserved charges in a covariant manner
by correlating the Abbott-Deser-Tekin (ADT) current~\cite{Abbott:1981ff,Abbott:1982jh,Deser:2002rt,Deser:2002jk,Senturk:2012yi} with the linearized Noether current.
Remarkably, the correspondence between these formalisms is established at the off-shell level, rendering it independent of the asymptotic behavior, in contrast to the original ADT method, which relies on asymptotic conditions.

In this paper, we aim to determine the conserved energy for the accelerating AdS black hole
using the quasilocal formalism~\cite{Kim:2013zha}.
Our study focuses on a slowly-accelerating black hole within anti-de Sitter (AdS) spacetime in order to admit only a single horizon~\cite{Podolsky:2002nk}.
This consideration ensures that a timelike Killing vector is uniquely determined, allowing the quasilocal energy to be well-defined.
Consequently, the energy expression, which turns out to be invariant along the radial direction, satisfies the thermodynamic first law of the black hole without introducing any additional thermodynamic variables.

The paper is organized as follows.
In Sec.~\ref{sec:Accelerating Anti-de Sitter black hole}, we find the Killing vector of the accelerating AdS black hole from the integrability condition of the thermodynamic first law.
In Sec.~\ref{sec:Conserved charge for the accelerating black hole}, we compute the quasilocal energy of the accelerating AdS black hole and obtain a new expression of the conserved energy.
We confirm the thermodynamic first law of the accelerating AdS black hole and discuss some differences from previous results.
The conclusion and discussion of our results are presented in Sec.~\ref{sec:Conclusion and discussion}.

\section{The Killing vector for the accelerating black hole}
\label{sec:Accelerating Anti-de Sitter black hole}
To obtain the quasilocal conserved energy of the accelerating AdS black hole, we start with the Einstein-Hilbert action defined by
\begin{equation}
	\label{eq:ads action}
	I = \frac{1}{16\pi G}\int \dd[4] x \sqrt{-g}\left( R + \frac{6}{\ell^2} \right),
\end{equation}
where $\ell$ denotes the AdS radius.
The solution for the accelerating AdS black hole can be expressed as
\begin{equation}
	\label{eq:acc ads metric}
	\dd s^2 = \frac{1}{\left( 1-A r \cos \theta \right)^2}\left[ -f(r)\dd t^2 + \frac{1}{f(r)}\dd r^2 + \frac{r^2}{g(\theta)}\dd \theta^2 + \frac{ g(\theta)r^2 \sin^2 \theta}{K^2} \dd \phi^2 \right],
\end{equation}
where the metric functions $f(r)$ and $g(\theta)$ are given as
\begin{equation}
	f(r) = (1-A^2r^2)\left( 1-\frac{2GM}{r} \right) + \frac{r^2}{\ell^2},\quad g(\theta) = 1 - 2AGM \cos\theta
\end{equation}
and the mass parameter $M$ and the acceleration parameter $A$ are
positive.
The AdS boundary is located at $r = \frac{1}{A \cos \theta}$ where $0\leq \theta < \frac{\pi}{2}$.
Since $\frac{1}{A \cos \theta}$ becomes negative in $\frac{\pi}{2} < \theta \leq \pi$, the metric \eqref{eq:acc ads metric} can only describe the part of the AdS boundary.
The deficit angles associated with the parameter $K$ are
calculated at the spacetime poles $\theta =0,\pi$ as
\begin{equation}
	\label{eq:acc sch deficit angle}
	\Delta \phi\big|_{\theta=0} = 2\pi \left( 1- \frac{1}{K}(1-2A G M) \right),\quad \Delta \phi\big|_{\theta=\pi} = 2\pi \left( 1- \frac{1}{K}(1+2A G M) \right).
\end{equation}
The particular choice such as $K = 1\pm 2AGM$ eliminates one of the deficit angles;
however, we assume $K$ to be an arbitrary constant,
independent of $M$ and $A$.
In addition, we restrict the parameter $A$ as $A < \frac{1}{\ell}$, referred as the slowly-accelerating black hole~\cite{Podolsky:2002nk}.
This restriction makes the black hole to admit a single horizon $r_h$, defined by $f(r_h)=0$, and then thereby a thermal temperature can be uniquely defined.
For $0< r_h < \frac{1}{A}$, the parameter $M$ is written as
\begin{equation}
	\label{eq:def rh}
	M = \frac{1}{2G}\left( r_h + \frac{r_h^3}{\ell^2(1-A^2r_h^2)} \right),
\end{equation}
where $f(r_h)=0$. In the limit $\ell \to \infty$, the horizon $r_h$ reduces to the Schwarzschild radius.

Next, let us consider the timelike Killing vector defined by $\xi = N\partial_t$, where $N$ is the dimensionless normalization factor which is assumed to depend on the parameters $r_h$, $A$, $\ell$, and $K$.
The black hole temperature is calculated as
\begin{equation}
	\label{eq:acc ads temp}
	T = \frac{N}{4\pi}\left( \frac{1-A^2r_h^2}{r_h} + \frac{r_h(3-A^2r_h^2)}{\ell^2(1-A^2r_h^2)} \right)
\end{equation}
and from Wald's entropy formula~\cite{Wald:1993nt,Wald:1999wa}, the black hole entropy can be expressed as
\begin{equation}
	\label{eq:acc ads entropy}
	S = -2\pi \int_{\mathcal{H}} \dd^2x \sqrt{|h|}\pdv{L}{R_{\mu\nu\rho\sigma}}\epsilon_{\mu\nu}\epsilon_{\rho\sigma} = \frac{\pi r_h^2}{KG(1-A^2r_h^2)}
\end{equation}
which is just the area of the black hole.

We will study the quasilocal conserved charge whose surface can be deformed freely without changing the charge as long as we do not pass through stress-energy sources. If we take the surface as the horizon of the black hole,
Wald's formulation~\cite{Wald:1993nt,Wald:1999wa} tells us that the charge should give
\begin{equation}
	\label{}
	\dd E = T \dd S.
\end{equation}
The above thermodynamic first law can always be
obtained by taking the surface of the quasilocal charge to be the horizon of the black hole. However, we should
check integrability to determine whether $E$ is a well-defined object or not.
Thus, the integrability condition is required to be
\begin{equation}
	\label{eq:integrability cond}
	0 = \dd(T\dd S) = \dd T \wedge \dd S,
\end{equation}
where we used the fact that $\dd^2 E = 0$.

The most general solution for the integrability condition \eqref{eq:integrability cond} is given by
\begin{equation}
	\label{eq:general normalization}
	N = \frac{\ell^2r_h(1-A^2r_h^2)}{\ell^2(1-A^2r_h^2)^2 + r_h^2(3-A^2r_h^2)}H(S),
\end{equation}
where $H(S)$ is an arbitrary differentiable function, and this normalization provides the black hole temperature \eqref{eq:acc ads temp} as $T = \frac{H(S)}{4\pi}$.
In particular, the function $H(S)$ is chosen as
\begin{equation}
	\label{eq:function H}
	H(S) =\sqrt{\frac{\pi}{G S}}= \frac{\sqrt{K(1-A^2r_h^2)}}{r_h}
\end{equation}
so that the Schwarzschild limit of the temperature can be obtained as  $T = \frac{1}{8\pi G M}$ when $\ell\to \infty$, $A\to0 $, and $ K\to1$.
Plugging Eq.~\eqref{eq:function H} into Eq.~\eqref{eq:general normalization}, we obtain
\begin{equation}
	\label{eq:norm factor}
	N(r_h, A, \ell, K) = \frac{\ell^2\left( K(1-A^2 r_h^2) \right)^{3/2}}{K\left( \ell^2(1-A^2r_h^2)^2 + r_h^2(3-A^2 r_h^2) \right)}.
\end{equation}
From a physical point of view, the normalization factor \eqref{eq:norm factor} is specifically chosen to reproduce the Schwarzschild limit of temperature.
\section{Conserved charge for the accelerating black hole}
\label{sec:Conserved charge for the accelerating black hole}
Now, the ADT charge corresponding to the Killing vector $\xi$, linearized with respect to an arbitrary background, is given by~\cite{Hyun:2014kfa,Hyun:2016dvt}
\begin{equation}
	\label{eq:ADT charge general}
	Q_{\rm ADT}(g \,;\,\xi,\delta g) = \frac{1}{16\pi G}\int_{B}\dd[2]x_{\mu\nu}\left( \delta K^{\mu\nu}(g\,;\,\xi) - K^{\mu\nu}(g\,;\,\delta\xi) - 2\xi^{[\mu}\Theta^{\nu]}(g\,;\,\delta g) \right),
\end{equation}
where $\dd[2]{x_{\mu\nu}} = \frac{1}{4} \epsilon_{\mu\nu \alpha\beta}\dd x^\alpha \wedge \dd x^\beta $ with $\epsilon_{tr\theta\phi}=-1$.
The Cauchy surface is defined as
\begin{equation}
	\label{eq:acc ads cauchy}
	\Sigma = \{(r ,\theta,\phi)\mid r_h <r < \rho,\ 0<\theta<\pi,\ 0<\phi<2\pi\},
\end{equation}
where $\rho$ is a fixed radius satisfying $\rho < \frac{1}{A}$ so that the Cauchy surface $\Sigma$ does not extend to the conformal infinity at $\theta=0$.
In Eq.~\eqref{eq:ADT charge general}, the partial boundary of the Cauchy surface $B$ is defined by
\begin{align}
	\label{eq:cauchy bdy}
	B &= \partial \Sigma_0 \cup \partial \Sigma_\rho \cup \partial \Sigma_\pi,\\
	\partial \Sigma_\rho &= \{(\rho,\theta,\phi)\mid \ 0<\theta<\pi,\ 0<\phi<2\pi\},\\
	\partial \Sigma_0 &= \{(r,0,\phi)\mid \ r_h<r<\rho,\ 0<\phi<2\pi\},\label{eq:0 cauchy bdy}\\
	\partial \Sigma_\pi &= \{(r,\pi,\phi)\mid \ r_h<r<\rho,\ 0<\phi<2\pi\}.\label{eq:pi cauchy bdy}
\end{align}
Note that when $\rho = r_h$, the boundary $B$ becomes the black hole horizon.
This means that the conserved charge corresponding to our surface defined in Eq.~\eqref{eq:cauchy bdy} should be independent of its radial parameter $\rho$.
Here, $K^{\mu\nu}$ is the off-shell Noether potential and $\Theta^\mu$ is the surface term generated from the metric variation of the action \eqref{eq:ads action}:
\begin{align}
	K^{\mu\nu}(g\,;\,\xi) &= 2\sqrt{-g}\nabla^{[\mu}\xi^{\nu]},\label{eq:noether potential}\\
	\Theta^\mu(g\,;\,\delta g) &= 2\sqrt{-g}\left( g^{\mu[\lambda}g^{\kappa]\nu}\nabla_{\kappa}\delta g_{\nu\lambda} \right),\label{eq:surface term}
\end{align}
where the metric variation of the metric is arbitrary at the off-shell level.
After the variation of metric, it is taken along the one-parameter path in the solution space in terms of $\lambda$, as  $0\leq\lambda\leq1$, by replacing $r_h$ with $ \lambda r_h$ in the solution.
Then, the Noether potential $K^{\mu\nu}$ is calculated as
\begin{align}
	K^{tr}&= \frac{ANr[(3-A^2r^2)(\ell^2\lambda r_h+(1-A^2\ell^2)\lambda^3r_h^3) -2r\ell^2(1-A^2\lambda^2r_h^2)]\sin\theta\cos\theta}{\ell^2(1-Ar\cos\theta)^3(1-A^2\lambda^2r_h^2)K}\nonumber\\
	&\quad-\frac{N[(1-A^2\ell^2)(2r^3(1-A^2\lambda^2r_h^2 ) + (1+A^2r^2)\lambda^3r_h^3) + (1+A^2r^2)\ell^2\lambda r_h] \sin\theta}{\ell^2(1-Ar\cos\theta)^3(1-A^2\lambda^2r_h^2)K},\label{eq:Noether potential tr} \\
	K^{t\theta} &= \frac{2ANr\left(\ell^2(1-A^2\lambda^2r_h^2) - A\lambda r_h(\ell^2 + (1-A^2\ell^2)\lambda^2r_h^2)\cos\theta \right)\sin^2\theta}{\ell^2(1-A^2\lambda^2r_h^2)\left( 1-Ar \cos\theta \right)^3K},\label{eq:Noether potential ttheta}
\end{align}
and the surface term $\Theta^\mu$ is also calculated as
\begin{align}
	\Theta^r &= \frac{r_h(3\lambda^2r_h^2-A^2\lambda^4r_h^4 + \ell^2(1-A^2\lambda^2r_h^2)^2)(1-3A^2r^2+Ar(3-A^2r^2)\cos\theta)\sin\theta}{\ell^2(1-A^2\lambda^2r_h^2)^2\left( 1-Ar \cos \theta \right)^3 K},\label{eq:surface term radial}\\
	\Theta^\theta &= \frac{Ar_h(3\lambda^2r_h^2 - A^2\lambda^4r_h^4+\ell^2(1-A^2\lambda^2r_h^2)^2)(2+6\cos 2\theta - Ar(9\cos\theta-\cos{3\theta}))}{4\ell^2(1-A^2\lambda^2r_h^2)^2\left( 1-Ar \cos \theta \right)^3 K}.\label{eq:surface term polar}
\end{align}
Inserting Eqs.~\eqref{eq:Noether potential tr}\,--\,\eqref{eq:surface term polar} into Eq.~\eqref{eq:ADT charge general}, one can obtain
\begin{align}
	Q_{\rm ADT}(g;\lambda)
	&= \frac{1}{16\pi G}\left( \int_{\partial \Sigma_{0,\lambda}} + \int_{\partial \Sigma_\rho}+ \int_{\partial \Sigma_{\pi,\lambda}} \right)\dd[2]x_{\mu\nu}\left( \delta_\lambda K^{\mu\nu}(\xi) - K^{\mu\nu}(\delta_\lambda\xi) - 2\xi^{[\mu}\Theta^{\nu]}(\delta_\lambda g) \right)\nonumber\\
	&= \frac{r_h}{2G\sqrt{K}(1-\lambda^2A^2r_h^2)^{3/2}},\label{eq:acc ads adt charge}
\end{align}
where the conical singularities on $\partial \Sigma_{0,\lambda}$ and $\partial \Sigma_{\pi,\lambda}$ are treated as boundaries:
\begin{align}
	\label{eq:cauchy bdy 0 pi}
	\partial \Sigma_{0,\lambda} &= \{(r,0,\phi)\mid \ \lambda r_h<r<\rho,\ 0<\phi<2\pi\},\\
	\partial \Sigma_{\pi,\lambda} &= \{(r,\pi,\phi)\mid \ \lambda r_h<r<\rho,\ 0<\phi<2\pi\}.
\end{align}
For the first and third integrations in Eq.~\eqref{eq:acc ads adt charge}, the angular component of the surface term $\Theta^{\theta}$ in Eq.~\eqref{eq:surface term} does not vanish along the azimuthal axis at $\theta = 0$ and $\theta=\pi$ so that the integration along these axes should be taken into account.

In the end, the quasilocal energy for the accelerating AdS black hole can be obtained as
\begin{equation}
	\label{eq:acc ads energy}
	E = Q(\xi) = \int_{0}^{1}Q_{\rm ADT}(g;\lambda)\dd \lambda = \frac{r_h}{2G\sqrt{K(1-A^2r_h^2)}}
\end{equation}
which is independent of the radial direction.
Note that the conserved energy \eqref{eq:acc ads energy} turns out to be $\sqrt{\frac{S}{4\pi G}}$.
The quasilocal energy \eqref{eq:acc ads energy} diverges positively in the limit $r_h \to \frac{1}{A}$, {\it i.e.}, $M \to \infty$ while in the limit $A\to 0$ it reduce to $\frac{r_h}{2G\sqrt{K}}$.
The behavior of the energy would be different from the previous results
in Refs.~\cite{Appels:2016uha,Appels:2017xoe,Gregory:2017ogk,Anabalon:2018ydc}.
In particular, the energy expressions in Refs.~\cite{Appels:2016uha,Appels:2017xoe,Gregory:2017ogk} are independent of the acceleration parameter $A$.
The energy expression in Ref.~\cite{Anabalon:2018ydc} is given by $\frac{M\sqrt{1-A^2\ell^2}}{K}$ which decreases as the acceleration parameter increases.

Next, we investigate the compatibility of the expression of the quasilocal energy \eqref{eq:acc ads energy} with the thermodynamic first law.
Using Eqs.~\eqref{eq:acc ads temp},
\eqref{eq:acc ads entropy}, and \eqref{eq:acc ads energy}, one can easily confirm that the energy
expression~\eqref{eq:acc ads energy} is completely
consistent with the simplest thermodynamic first law:
\begin{equation}
	\label{eq:acc ads 1st law w/o A lambda}
	  \dd E = T\dd S.
\end{equation}
This result differs from the form of the first law of thermodynamics presented in Refs.~\cite{Appels:2016uha,Appels:2017xoe,Gregory:2017ogk,Anabalon:2018ydc}, where terms proportional to the variation of string tensions appear.
\section{Conclusion and discussion}
\label{sec:Conclusion and discussion}
In this paper, we have obtained the conserved quasilocal energy of the accelerating AdS black hole
by using the quasilocal formalism and investigated the thermodynamic first law in connection with the new energy expression.
The expression of the energy eventually satisfies the standard form of the thermodynamic first law
without any modifications.

Presently, one might wonder why the energy expression~\eqref{eq:acc ads energy} is different from that obtained in Ref.~\cite{Appels:2017xoe}.
The origin of the discord seems to emanate from the difference of definitions
between the quasilocal ``Komar'' formalism~\cite{Komar:1958wp}
and the quasilocal ``ADT'' formalism~\cite{Kim:2013zha}, and the normalization choice of the Killing vector.
In Ref.~\cite{Appels:2017xoe}, the quasilocal expression for the Komar energy of the accelerating black hole is obtained by
$Q_{\rm Komar} = -\frac{1}{4\pi G}\int_{S^2}\dd[2]{x}_{\mu\nu}\nabla^\mu \xi^\nu = \frac{M}{K}$,
where the timelike Killing vector is chosen as $\xi= \partial_t$, and $S^2$ is located in the finite region of the spacetime.
On the other hand,  the quasilocal formalism we have adopted is based on the amalgamation between the ADT method~\cite{Abbott:1981ff,Abbott:1982jh,Deser:2002rt,Deser:2002jk,Senturk:2012yi} and the covariant phase space method~\cite{Iyer:1994ys,Wald:1999wa}, both of which represent covariant generalizations of the ADM formalism~\cite{Arnowitt:1962hi}.
The discrepancy of our energy expression to other ones appears due to the additional contribution
of the boundary term, related to the conical singularities.
In summary, one of the key observations in our paper is that loci, $\theta=0$ and $\theta=\pi$ where the conical singularities are located,
must be regarded as one of boundaries in the computation of quasilocal charges.

The conserved charge can also be discussed using the covariant phase space method, in which the Hamiltonian can be derived directly from the action~\cite{Kim:2023ncn}.
For the contributions of the conical singularity, the divergent polar component of the surface term \eqref{eq:surface term polar} was renormalized at conformal infinity by introducing counterterms, and then, 
the thermodynamic length and the string tension were obtained independent of the temperature and the entropy.
Hence, the terms proportional to the variation of the string tension appeared in the thermodynamic first law.
These terms may be rewritten in terms of closed forms and thus the energy variation may be effectively redefined
in order to obtain the thermodynamic first law without tension terms. However, the conserved energy would be different from
our energy expression because the choice of the normalization of the Killing vector is different.

Finally, for the Smarr relation~\cite{Smarr:1972kt}, we can use the scaling method described in Ref.~\cite{Kubiznak:2016qmn}.
From the energy expression \eqref{eq:acc ads energy}, one can obtain the scaling relations:
$r_h \to \alpha r_h$, $A \to \alpha^{-1} A$, $E \to \alpha E$, $S \to \alpha^2 S$.
Euler's homogeneous function theorem, along with  the energy \eqref{eq:acc ads energy}, tells us that $E = 2S\left( \pdv{E}{S} \right) = 2TS,$
which is the Smarr relation for the accelerating AdS black hole.

\acknowledgments
We would like to thank Gabriel Arenas-Henriquez, Dario Martelli, Adam Ball, Aaron Poole, and Jeong-Hyuck Park for valuable comments and discussions.
This work was supported by Basic Science Research Program through the National Research Foundation of Korea(NRF) funded by the Ministry of Education through the Center for Quantum Spacetime (CQUeST) of Sogang University (NRF-2020R1A6A1A03047877).
WTK was supported in part by the National Research Foundation of Korea(NRF) grant funded by the Korea government(MSIT) (No.\! NRF-2022R1A2C1002894) and SHY was supported in part  by the National Research Foundation of Korea(NRF) grant funded by the Korea government(MSIT) (No.\! NRF-2021R1A2C1003644).

\bibliographystyle{JHEP}       

\bibliography{reference}

\end{document}